\documentclass[prb,twocolumn,showpacs,aps]{revtex4}

\usepackage{graphicx}%
\usepackage{dcolumn}
\usepackage{amsmath}

\begin{document}


\title{Acoustic properties of colloidal crystals}

\author{I.E.~Psarobas}

\thanks{Also at the Section of Solid State Physics of the University of Athens,
Panepistimioupolis, GR-157 84, Athens, Greece.}

\email{ipsarob@cc.uoa.gr} \homepage{http://www.uoa.gr/~vyannop}

\affiliation{Department of Physics, National Technical University
of Athens\\ Zografou Campus, GR-157 73, Athens, Greece}
\date{\today}
\author{R.~Sainidou}
\affiliation{Section of Solid State Physics, University of Athens,
Panepistimioupolis,\\ GR-157 84, Athens, Greece}
\author{N.~Stefanou}
\affiliation{Section of Solid State Physics, University of Athens,
Panepistimioupolis,\\ GR-157 84, Athens, Greece}
\author{A.~Modinos}
\affiliation{Department of Physics, National Technical University
of Athens\\ Zografou Campus, GR-157 73, Athens, Greece}
\begin{abstract}
We present a systematic study of the frequency band structure of
acoustic waves in crystals consisting of nonoverlapping solid
spheres in a fluid. We consider colloidal crystals consisting of
polystyrene spheres in water, and an opal consisting of
close-packed silica spheres in air. The opal exhibits an
omnidirectional frequency gap of considerable width; the
colloidal crystals do not. The physical origin of the bands are
discussed for each case in some detail. We present also results
on the transmittance of finite slabs of the above crystals.
\end{abstract}

\maketitle
\section{Introduction}
\label{intro} Colloidal suspensions of monodisperse spherical
particles, with a diameter between 1 nm and 1 $\mu$m, in liquids
or gases self assemble onto three-dimensional simple lattices
whose lattice parameter can be easily tailored, providing new
opportunities for fundamental as well as for applied research
(see, e.g., Ref.~\onlinecite{colpbg} and references therein). The
optical properties of these colloidal crystals are being
investigated by many research groups because they seem to be good
photonic band-gap materials; but they have interesting acoustic
properties as well.~\cite{Sprik,Psheng1,Psheng2,Psheng3} Moreover,
experiments relating to the reflection, transmission and
absorption of ultrasonic waves by colloidal crystals can be very
useful for the characterization of such systems, provided the
means exist for a proper theoretical analysis of the experimental
data. Finally, theoretical results for colloidal crystals offer a
starting point for the understanding, at a semiquantitative
level, of the propagation of acoustic waves in corresponding
random media consisting of monodispersed spherical particles in a
fluid.~\cite{Psheng1,Psheng2,Psheng3}\begin{figure}
\centerline{\includegraphics*[height=5cm]{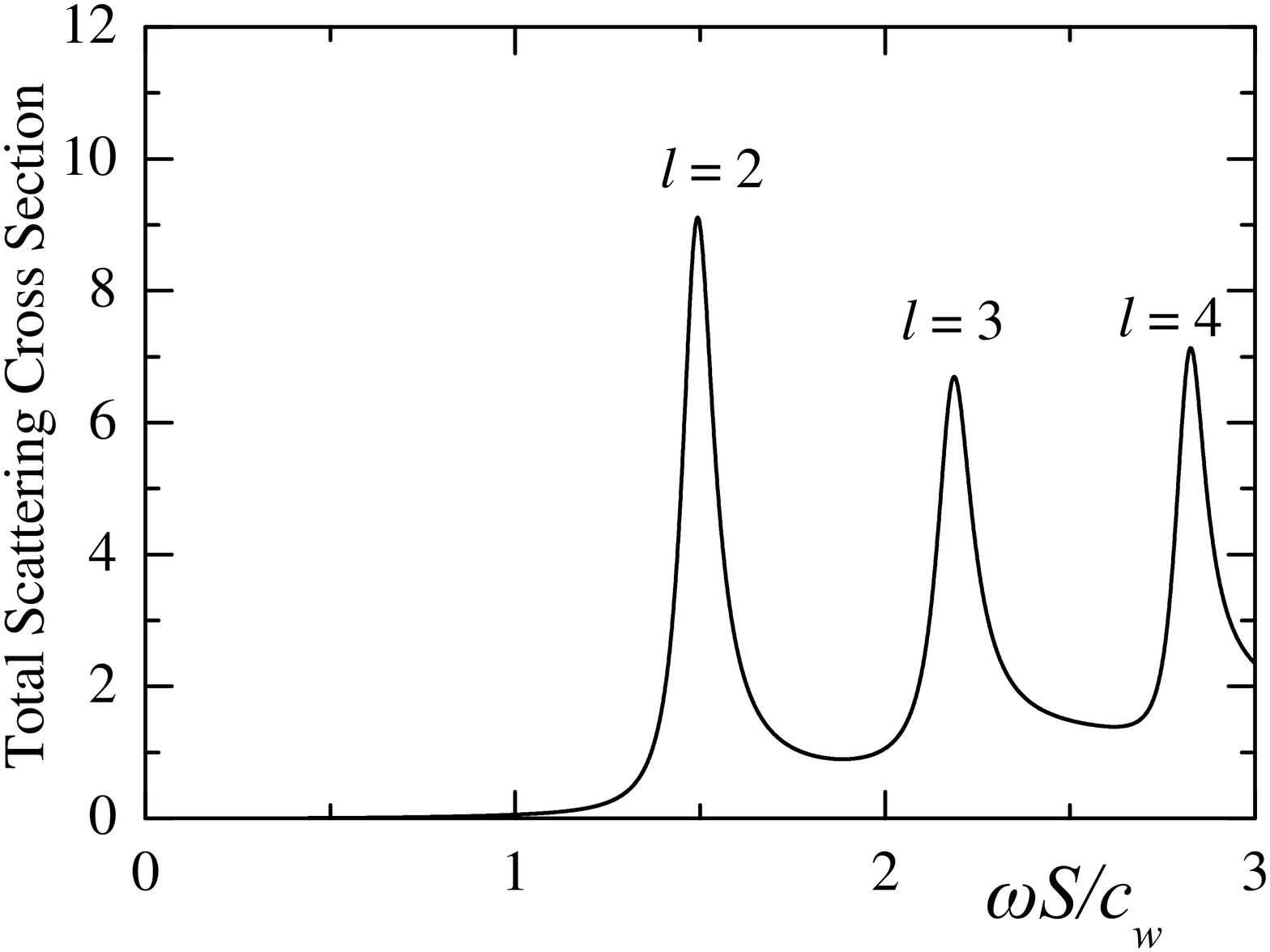}}
\caption{Scattering of a plane acoustic wave by a polystyrene
sphere in water: total scattering cross section.} \label{fig1}
\end{figure}

The vibration modes (normal modes of the elastic field) of a
phononic crystal, by which we mean a composite material whose
density $\rho({\bf r})$ and Lam\'e coefficients $\lambda({\bf
r})$ and $\mu({\bf r})$ vary periodically in space, are Bloch
waves with a corresponding frequency band structure which is
analogous to that of electrons in ordinary crystals and
electromagnetic waves in photonic
crystals.~\cite{Sig,Kushw,Kushw2,Kafesaki,Sig2,ipsarob1,CT1} With
an appropriate choice of the parameters involved one may obtain
phononic crystals with absolute (omnidirectional) frequency gaps
(phononic gaps) in selected regions of frequency. An elastic
wave, whose frequency lies within an absolute frequency gap of a
phononic crystal, incident on a slab of the crystal of certain
thickness will be reflected by it, the slab operating as a perfect
nonabsorbing mirror of elastic waves in the frequency region of
the gap.~\cite{Nature,CT2} It may be possible, for example, to
construct in this way vibration-free cavities which might be very
useful in high-precision mechanical systems operating in a given
frequency range.

In this paper we investigate in detail the acoustic properties of
fcc colloidal crystals of polystyrene spheres in water and of
close-packed silica spheres in air (opals). Given that these
systems have a characteristic length scale of the order of 1
$\mu$m, they should exhibit interesting acoustic properties at
ultrasonic frequencies of a few hundred MHz to a few GHz. It
should be noted, however, that our results apply to different
regions of frequency of the acoustic field provided that the size
of the spheres and of the unit cell are scaled accordingly, and
provided the elastic coefficients can be taken as constants
independent of frequency over the said regions. The calculations
of the frequency band structure and of the transmission
coefficient of acoustic waves through a slab of the material were
done using the layer-multiple-scattering formalism we have
developed for this purpose.~\cite{ipsarob1} A formalism along the
same lines has also been published by Liu {\em et
al.}.~\cite{CT1} We have already demonstrated the efficiency of
this method in relation to solid-solid
composites.~\cite{ipsarob2} The present paper shows that the
method applies equally well to phononic crystals consisting of
nonoverlapping solid spheres in a fluid host; it appears that the
plane-wave method for calculating the frequency band structure of
such systems has convergence problems.~\cite{Kafesaki} Besides,
we shall pay particular attention to the physical origin of the
different modes of the acoustic field in the systems under
consideration.
\section{Polystyrene spheres in water}
\label{poly} We consider a model colloidal crystal consisting of
polystyrene spheres in water. The mass density and the sound
velocities for polystyrene are: $\rho_p=1050\  {\rm kg/m^3}$,
$c_{lp}=2400 \ {\rm m/s}$, $c_{tp}=1150 \ {\rm m/s}$. For water
we have: $\rho_w=1000 \ {\rm kg/m^3}$, $c_{w}=1480 \ {\rm m/s}$.
We begin by considering the scattering of a harmonic plane
acoustic wave (a longitudinal wave) of angular frequency
$\omega$, by a single polystyrene sphere of radius $S$ in water;
the water extends over all space. The incident plane wave can be
written as a sum of spherical waves associated with the spherical
harmonics $Y_{lm}$, where $l=0,1,2,\cdots$ and
$m=-l,\cdots,0,\cdots,l$, as usual.~\cite{ipsarob1}\begin{figure}
\centerline{\includegraphics*[height=6cm]{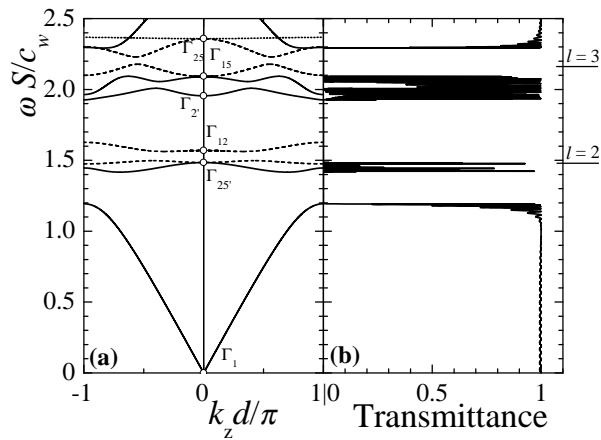}} \caption{(a)
The phononic frequency band structure normal to the (111) surface
of a fcc crystal of polystyrene spheres in water with $f=30\%$.
The solid/dotted/dashed lines refer to bands of
$\Lambda_1$/$\Lambda_2$/$\Lambda_3$ symmetry, respectively. (b)
Transmittance of an acoustic wave incident normally on a slab of
the above crystal consisting of 32 planes of spheres parallel to
the (111) surface. The positions of the first two resonant modes
of a single polystyrene sphere in water are indicated in the
margin.} \label{fig2}
\end{figure} A spherical
wave of given $(l,m)$ scatters independently of the others, of
different ($l,m$), because of the spherical symmetry of the
scatterer; therefore the total scattering cross section is the
sum of partial $(l,m)$ cross sections, with $l$ up to a maximum
$l_{max}$ depending on the size parameter, $\omega S/c_w$, of the
sphere. We assume that waves with $l>l_{max}$ do not scatter from
the sphere and do not contribute to the total scattering cross
section. Of course partial cross sections of different $m$ (but of
the same $l$) are equal because of the spherical symmetry of the
system. In Fig.~\ref{fig1} we show the total scattering cross
section for a plane wave scattered by a polystyrene sphere in
water as a function of $\omega S/c_w$. The peaks correspond to
resonant modes of the acoustic field about the sphere with an
angular distribution of the displacement field at the surface of
the sphere determined by the spherical waves which contribute to
these modes. We find that the resonance at $\omega S/c_{w}=1.49$
is an $l=2$ resonance: the displacement field associated with it
is made up almost entirely (98\%) from $l=2$ spherical waves; the
one at $\omega S/c_{w}=2.19$ is an $l=3$ resonance: the
corresponding displacement field is made up mostly (90\%) from
$l=3$ spherical waves; finally the field associated with the
resonance at $\omega S/c_{w}=2.83$ is made up mostly from $l=4$
waves (65\%) and $l=1$ waves (22\%). And we remember that there
will be ($2l+1$) resonant modes of the displacement field
corresponding to $m=-l,\cdots,0,\cdots,l$ of the same frequency
$\omega_l$, i.e. $\omega_l$ is ($2l+1$)-degenerate.

We now consider a fcc crystal, with lattice constant $a$, of
nonoverlapping polystyrene spheres in water. \begin{figure}
\centerline{\includegraphics*[height=6cm]{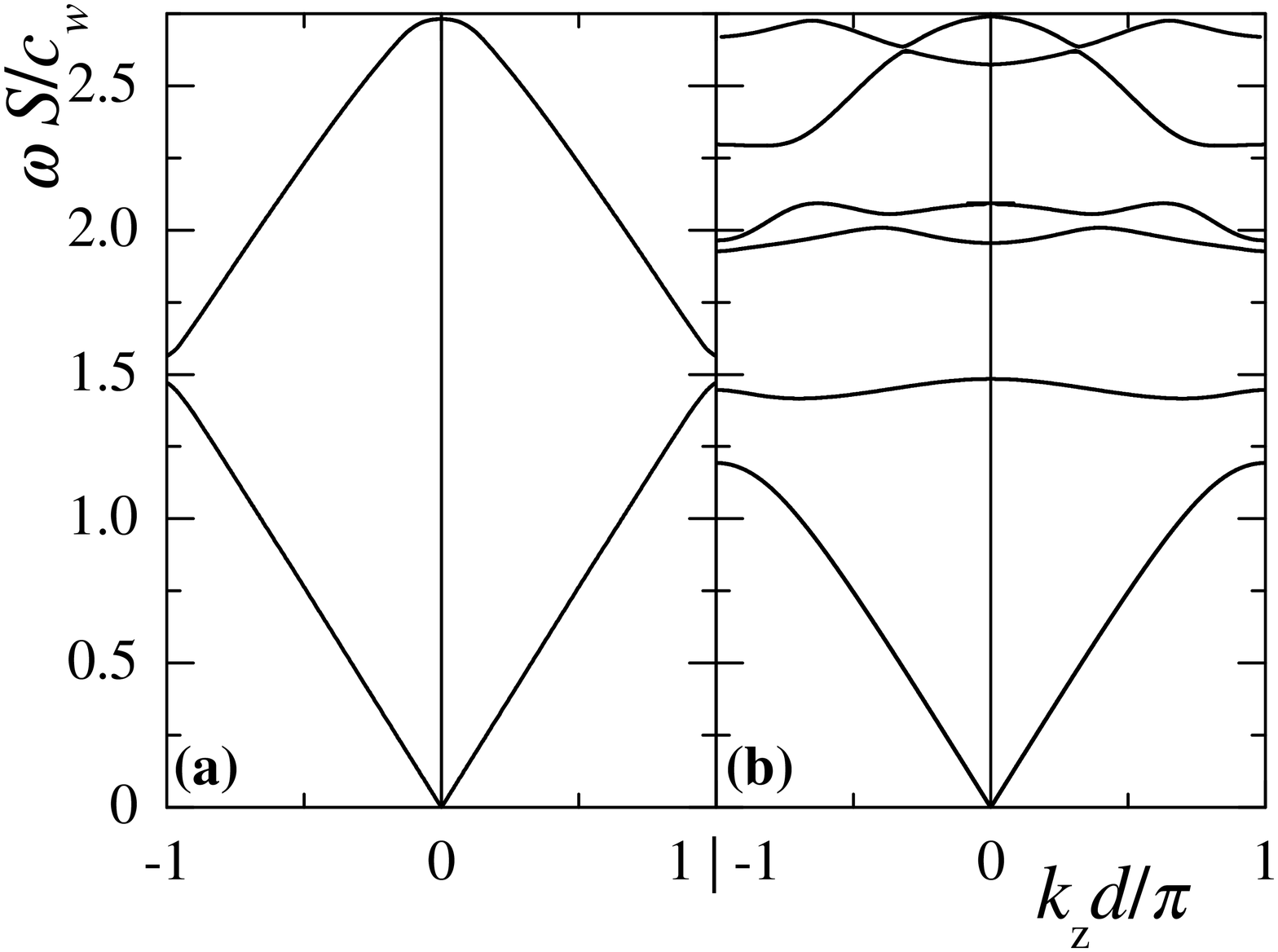}} \caption{The
phononic frequency band structure normal to the (111) surface of
a fcc crystal of polystyrene spheres in water with $f=30\%$ (only
bands of $\Lambda_1$ symmetry are shown), calculated (a): with
$l_{max}=1$ and (b): with $l_{max}=6$.} \label{fig3}
\end{figure}We view the crystal
as a sequence of (111) planes of spheres. The spheres of a plane
are arranged on a hexagonal lattice of lattice constant
$a_0=a\sqrt{2}/2$, defined by the primitive vectors ${\bf
a}_1=a_0(1,0)$ and ${\bf a}_2=a_0(1/2,\sqrt{3}/2)$, in the $xy$
plane (plane of the spheres). The ($n+1$)th plane along the $z$
axis is obtained from the $n$th plane by a primitive translation,
${\bf a}_3$, of the crystal. The planes are separated along the
$z$ direction by a distance $d=a\sqrt{3}/3$.

Fig.~\ref{fig2}(a) shows the frequency band structure of acoustic
waves in this crystal normal to the (111) surface, obtained by
the method of Ref.~\onlinecite{ipsarob1}, when the fractional
volume occupied by the spheres is $f=30\%$. In this case, the
component of the reduced wavevector within the surface Brillouin
zone (SBZ) of the fcc (111) surface, ${\bf k}_{\|}$, equals zero.
The symmetry of the bands $k_z(\omega ; {\bf k}_{\|}={\bf 0})$ is
determined by the symmetry group for this direction (the $C_{3v}$
group) and the corresponding propagating modes (Bloch waves) of
the acoustic field belong to the different irreducible
representations of this group, denoted by $\Lambda_1$,
$\Lambda_2$ and $\Lambda_3$ (see, e.g.,
Ref.~\onlinecite{GroupTheory}); $\Lambda_1$ and $\Lambda_2$ are
one-dimensional: the corresponding frequency bands are
non-degenerate, and $\Lambda_3$ is two-dimensional: the
corresponding frequency bands are doubly degenerate. At the
$\Gamma$ point (${\bf k}_{\|}={\bf 0}$, $k_z=0$) the modes of the
acoustic field belong to the irreducible representations of the
$O_h$ group associated with this point, as shown in
Fig.~\ref{fig2}(a). We see that the frequency bands of
Fig.~\ref{fig2}(a) appear in pairs: $k_z(\omega ; {\bf
k}_{\|}={\bf 0})$, $-k_z(\omega ; {\bf k}_{\|}={\bf 0})$.
However, this symmetry property is not valid for an arbitrary
value of ${\bf k}_{\|}$, because the crystal under consideration,
described by the $O_h$ group, does not have a plane of mirror
symmetry parallel to the (111) surface which would transform
$(k_x,k_y,k_z)$ to $(k_x,k_y,-k_z)$. The above symmetry property
holds only for ${\bf k}_{\|}$ on the $x$ axis [${\bf
k}_{\|}=(k_x,0)$], because a rotation through $\pi$ about the $x$
axis is a symmetry operation of $O_h$ and it transforms
$(k_x,0,k_z)$ to $(k_x,0,-k_z)$.

The (longitudinal) wavefield, of given $\omega$ and ${\bf
k}_{\|}$, in the host region between the $n$th and the ($n+1$)th
planes of spheres, can be expanded into plane waves propagating
(or decaying) to the left and to the right as follows
\begin{widetext}
\begin{eqnarray}
{\bf u}(\omega;{\bf k}_{\|})=  \sum_{{\mathbf{g}}} \biggl\{{\bf
u}_{{\mathbf{g}}n}^{+}
\exp\left[i\,{\mathbf{K}}_{{\mathbf{g}}l}^{+} \cdot
({\mathbf{r}}-{\mathbf{A}}_n)\right]+\ {\bf u}_{{\mathbf{g}}n
}^{-} \exp\left[i\,{\mathbf{K}}_{{\mathbf{g}}l}^{-} \cdot
({\mathbf{r}}-{\mathbf{A}}_n)\right]\biggr\}\;, \label{eq:unn+1}
\end{eqnarray}
\end{widetext}
with
\begin{equation}
{\bf K}_{{\bf g}}^{\pm}= \left({\bf k}_{\|}+{\bf g},\,\pm
\left[(\omega /c_w)^2-({\bf k}_{\|}+{\bf
g})^2\right]^{1/2}\right)\;, \label{eq:Kg}
\end{equation}
where ${\bf g}$ are the two-dimensional reciprocal vectors
corresponding to the lattice defined by ${\bf a}_1$ and ${\bf
a}_2$ above, and ${\bf A}_n$ is a point between the $n$th and
$(n+1)$th planes. A generalized Bloch wave satisfies the equation
\begin{equation}
{\bf u}_{{\mathbf{g}}n+1}^{\pm}= \exp\left(i{\mathbf{k}}
\cdot{\mathbf{a}}_{3}\right) {\bf u}_{{\mathbf{g}}n}^{\pm}\,,
\label{eq:blochwave}
\end{equation}
where ${\bf a}_3={\bf A}_{n+1}-{\bf A}_n$ and
${\mathbf{k}}=\left({\mathbf{k}}_{\|},k_{z}(\omega;{\mathbf{k}}_{\|})
\right)$. We note that, although both longitudinal and transverse
modes of the elastic field are considered within the (solid)
spheres, only longitudinal modes exist in the (fluid) host region
in the absence of viscosity. Therefore, in a binary composite of
nonoverlapping solid spheres in a fluid host, where the solid
component does not form a continuous network, there cannot be
propagating transverse waves. Consequently, at the $\Gamma$ point
(${\bf k}_{\|}={\bf 0},$ $k_z=0$) the dispersion curves
[Fig.~\ref{fig2}(a) and also
Figs.~\ref{fig3},~\ref{fig5},~and~\ref{fig7}(a) below], show only
one (longitudinal) branch starting from zero frequency, instead of
the three (corresponding to both longitudinal and transverse
modes) appearing in solid-solid
composites.~\cite{Sprik,Sig,Sig2,ipsarob1,CT1,CT2,ipsarob2}

One can easily show that when ${\bf k}_{\|}={\bf 0}$, the ${\bf
g}={\bf 0}$ component of the wavefield, described by
Eq.~(\ref{eq:unn+1}), vanishes (${\bf u}_{{\mathbf{g}={\bf 0}}
}^{\pm}={\bf 0}$) for the modes of the $\Lambda_2$ or the
$\Lambda_3$ symmetry; only the $\Lambda_1$ symmetry allows a
non-vanishing ${\bf g}={\bf 0}$ component of the wavefield. Now
assume that we have a slab of the crystal of finite thickness,
i.e. $N$ fcc (111) planes of polystyrene spheres; between the
spheres and to the left and right of the slab, extending to
infinity, we have water. There, and for the frequency range which
interests us here, the acoustic wavefield has only the one
component ${\bf g}={\bf 0}$. And therefore the external field
couples with the field inside the slab essentially only through
the ${\bf g}={\bf 0}$ component of the latter. Therefore an
acoustic wave of given frequency, incident normally on the (111)
slab of the crystal, will excite essentially only $\Lambda_1$
modes of the crystal. If such a band does not exist at the given
frequency the wave will be totally reflected by the slab. There
will be no transmitted wave. The reader will see that this is
indeed the case in the present instance: we show in
Fig.~\ref{fig2}(b), opposite the frequency band structure, the
transmission coefficient of an acoustic plane wave incident
normally on a slab of the crystal consisting of 32 planes of
polystyrene spheres in water. The transmission coefficient
opposite the $\Lambda_1$ bands exhibits the well-known
Fabry-Perot-like oscillations due to multiple scattering between
the surfaces of the slab; elsewhere it practically vanishes, and
this includes regions of frequency where only bands of the
$\Lambda_2$ and/or $\Lambda_3$ symmetry exist. Of course the above
argument holds for normal incidence; off this direction the
symmetry argument does not apply, and there will be some
transmission if there are any propagating modes of the acoustic
field at the given frequency (see below, Fig.~\ref{fig4}).

From a practical point of view, once the frequency band structure
and, if required, the transmission coefficient have been
calculated, one has all that is necessary for a comparison with
relevant experimental data. However, it is worthwhile to look at
the physics behind the band structure of Fig.~\ref{fig2}. We note
that the five narrow bands about $\omega S/c_w\approx1.5$ (we
remember that the $\Lambda_3$ band is doubly degenerate) derive
from the $l=2$ resonances on the individual spheres, which
interact weakly between them. The five-fold degeneracy of the
resonance of the single sphere is split by this interaction in
accordance with the lower (cubic) symmetry of the crystal field.
Similarly the seven narrow bands about $\omega S/c_w\approx2.2$
derive from the $l=3$ resonances on the individual spheres. We
observe the small hybridization gaps opening up about $\omega
S/c_w\approx1.5$ and about $\omega S/c_w\approx2.2$, between
bands of the same symmetry.\begin{figure}
\centerline{\includegraphics*[height=5cm]{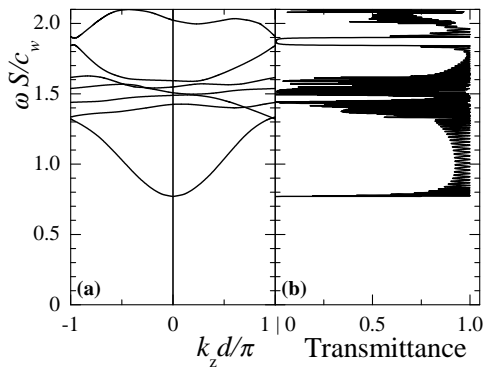}} \caption{(a)
The phononic frequency band structure associated with the (111)
surface of a fcc crystal of polystyrene spheres in water with
$f=30\%$, for ${\bf k}_{\|}=\frac{2\pi}{a_0}(0.3,0.1)$. (b)
Transmittance of an acoustic wave incident with ${\bf
k}_{\|}=\frac{2\pi}{a_0}(0.3,0.1)$ on a slab of 32 fcc (111)
planes of polystyrene spheres in water.} \label{fig4}
\end{figure}

There is also a mode of propagation of acoustic waves
corresponding to almost free propagation in an effective
homogeneous medium. In the absence of the resonances associated
with the spheres, this mode of propagation, which has the
$\Lambda_1$ symmetry, would dominate the frequency band structure
of the acoustic field as shown in Fig.~\ref{fig3}(a). It was
calculated by suppressing the $l>1$ resonances, i.e. cutting off
the $l>1$ spherical waves, which give rise to these resonances,
from the spherical-wave expansion of the wavefield (see
Ref.~\onlinecite{ipsarob1}). In the long-wavelength limit
($\omega \rightarrow 0$) we obtain a linear dispersion curve, the
slope of which gives an effective velocity of sound for the
composite medium, $\bar{c}=1589 \ {\rm m/s}$, which is in very
good agreement with the result, $\bar{c}=1566 \ {\rm m/s}$, of the
effective-medium approximation.~\cite{Gaunaurd} The small gap
about $\omega S/c_w\approx1.5$ in Fig.~\ref{fig3}(a), is a Bragg
gap; it is analogous to the small gaps one obtains at the edges
of the Brillouin zone in the electronic band structure of
free-electron-like metals. When the resonances on the spheres are
allowed in, one obtains the band structure shown in
Fig.~\ref{fig3}(b), with apparent hybridization gaps between the
continuum band [of Fig.~\ref{fig3}(a)] and resonance bands of the
same symmetry ($\Lambda_1$).

We believe that the disorder that is naturally there in a
colloidal solution, does not eradicate the essential
characteristics of the acoustic modes as calculated here, which
are determined by the local environment about a sphere and less so
by long range order. We would expect the fine features associated
with narrow bands to be smoothed out by disorder, but bands of
modes separated by gaps could remain. This would be in accordance
with the results of Brillouin-scattering experiments on
disordered colloidal suspensions of monodispersed
polymethylmethacrylate (PMMA) spheres in transparent oil. It has
been shown that in these colloidal suspensions different
longitudinal modes of propagation of the acoustic field exist
which are separated by frequency gaps about the resonance
frequencies of an individual PMMA
sphere,~\cite{Psheng1,Psheng2,Psheng3} which agrees, at a
semiquantitative level, with the hybridization-induced gaps
discussed above.

In Fig.~\ref{fig4}(a) we show an example of the band structure
for a ${\bf k}_{\|}\neq {\bf 0}$, and next to it, in
Fig.~\ref{fig4}(b), the transmission coefficient of an acoustic
wave, with the same ${\bf k}_{\|}$, incident on a slab of the
material consisting of 32 fcc (111) planes of polystyrene spheres.
The thing to note is that all bands are active in this case,
although the transmission coefficient is not always unity
[compare with Fig.~\ref{fig2}(a)]. We remember that the incident
wave cannot have a frequency smaller than
$\omega_{inf}=c_w\left|{\bf k}_{\|}\right|$. Finally, it can be
seen that now the frequency bands do not have the reflection
symmetry found in the case of ${\bf k}_{\|}={\bf 0}$
[Fig.~\ref{fig2}(a)], for the reasons given above.

We calculated the frequency band structure for a sufficient
number of ${\bf k}_{\|}$ points within the SBZ. There is a
near-omnidirectional gap about $\omega S/c_w\approx1.3$; we tried
to turn this gap into a proper omnidirectional gap by changing
some of the parameters, but we did not succeed.\begin{figure}
\centerline{\includegraphics*[height=5cm]{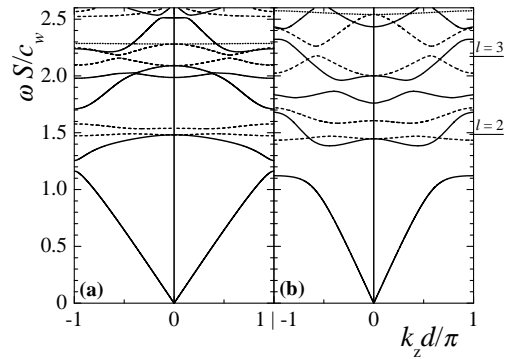}} \caption{The
phononic frequency band structure normal to the (111) surface of
a fcc crystal of polystyrene spheres in water with (a): $f=20\%$
and (b): $f=50\%$. The solid/dotted/dashed lines refer to bands
of $\Lambda_1$/$\Lambda_2$/$\Lambda_3$ symmetry, respectively.
The positions of the first two resonant modes of a single
polystyrene sphere in water are indicated in the
margin.}\label{fig5}
\end{figure}

In Fig.~\ref{fig5} we look at the dependence of the frequency band
structure on the fractional volume occupied by the spheres. We
see in particular that the width of the resonance bands increases
with $f$, apparently because the spatial overlap of the wavefield
associated with resonances on neighboring spheres increases with
$f$.
\section{Opals}
\label{opals} In this section we consider a phononic crystal, an
artificial opal, consisting of two media (silica spheres in air)
with density and velocity contrasts much higher than was the case
in the crystal (polystyrene spheres in water) studied in the
preceding section. Opalescent structures (usually referred to as
synthetic or artificial opals) can be obtained from monodispersed
silica colloids, e.g. by sedimentation in the gravitational
field. In this way, fcc arrays of silica microspheres in air,
near the close-packing density ($\sim 74 \%$), similar to a
natural opal, have been obtained.~\cite{colpbg} Here we present
some results on the acoustic properties of such an opal. The mass
density and the sound velocities for silica are: $\rho_s=2200\
{\rm kg/m^3}$, $c_{ls}=5970 \ {\rm m/s}$, $c_{ts}=3760 \ {\rm
m/s}$. For air we have: $\rho_a=1.23 \ {\rm kg/m^3}$, $c_a=340 \
{\rm m/s}$.

\begin{figure}
\centerline{\includegraphics*[height=5cm]{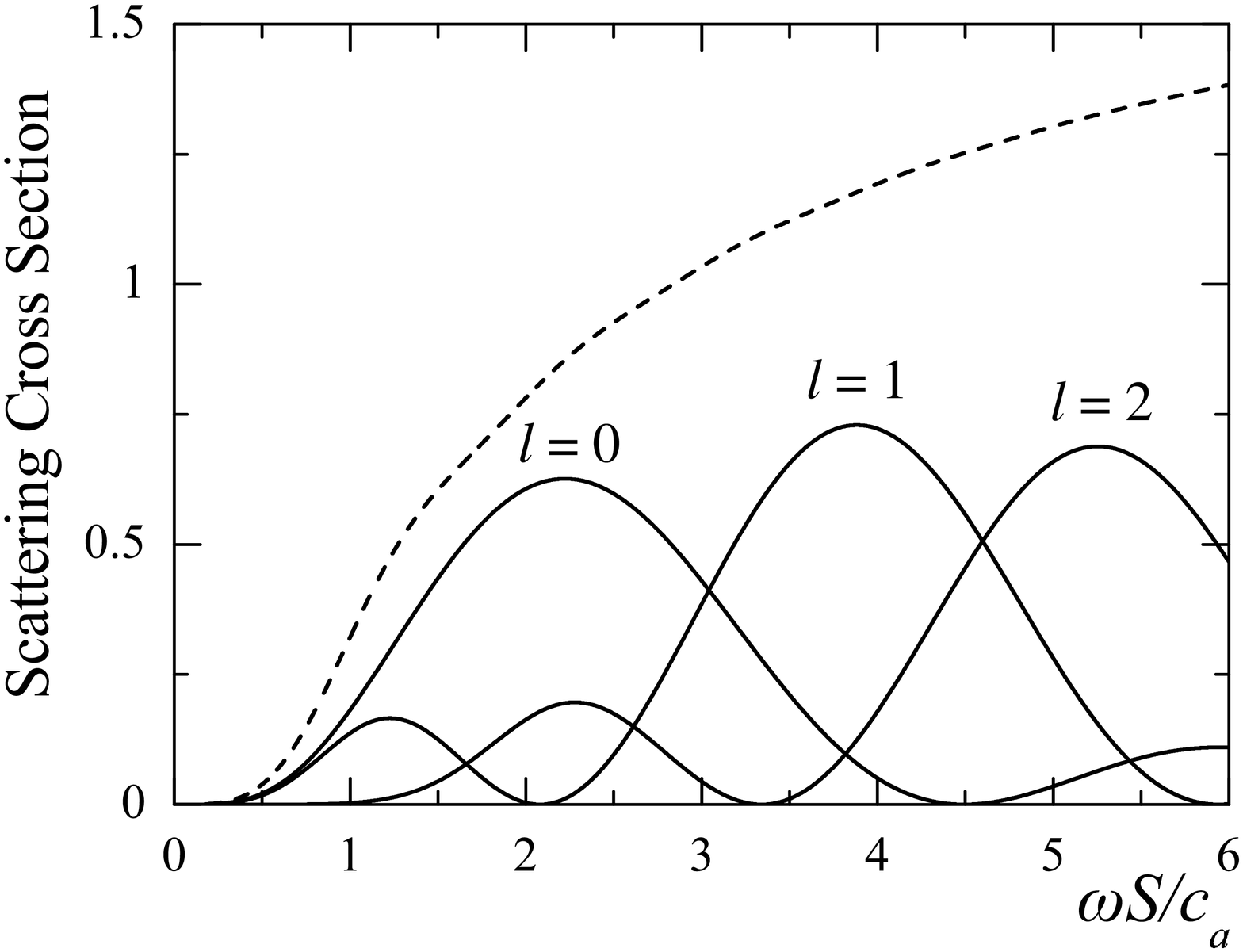}}
\caption{Scattering of a plane acoustic wave by a silica sphere
in air. Solid lines: partial scattering cross sections
corresponding to $l=0,1,2$. Dashed line: total scattering cross
section.}\label{fig6}
\end{figure}

Proceeding as in Section~\ref{poly}, we first consider the
scattering of a harmonic plane acoustic wave, of angular
frequency $\omega$, by a single silica sphere of radius $S$ in
air. In this case the total scattering cross section is a slowly
increasing function of $\omega S/c_a$ (see
Fig.~\ref{fig6}).
However, the partial cross sections, corresponding to different
values of $l$, exhibit well defined resonance peaks, of some
width, as can be seen from Fig.~\ref{fig6}. At $\omega S/c_a=2.22$
we obtain an $l=0$ resonance, at $\omega S/c_a=3.88$ an $l=1$
resonance, and at $\omega S/c_a=5.25$ an $l=2$ resonance.

\begin{figure}
\centerline{\includegraphics*[height=6cm]{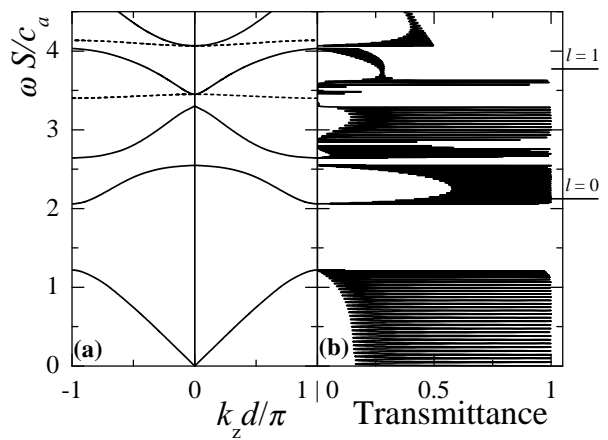}} \caption{(a)
The phononic frequency band structure normal to the (111) surface
of a fcc crystal of silica spheres in air with $f=74\%$. The
solid and dashed lines refer to bands of $\Lambda_1$ and
$\Lambda_3$ symmetry, respectively. (b) Transmittance of an
acoustic wave incident normally on a slab of the above crystal
consisting of 32 planes of spheres parallel to the (111) surface.
The positions of the first two resonant modes of a single silica
sphere in air are indicated in the margin.}\label{fig7}
\end{figure}

We consider then an fcc crystal of nonoverlapping silica spheres
in air with $f=74\%$ and view it as a sequence of (111) planes of
spheres. Fig.~\ref{fig7}(a) shows the frequency band structure of
the infinite crystal, for ${\bf k}_{\|}={\bf 0}$. The physical
origin of the bands shown in Fig.~\ref{fig7}(a), other than the
flat band about $\omega S/c_a=3.426$, can be understood in much
the same way as for the bands of Fig.~\ref{fig2}(a). In the
absence of hybridization we would have, extending practically over
the entire frequency range, a mode corresponding to almost free
propagation in an effective homogeneous medium, with Bragg gaps
opening up about $\omega S/c_a\approx1.5$ at the edges of the
Brillouin zone and about $\omega S/c_a\approx3.5$ at the center of
the Brillouin zone; and we would have also resonance bands
developing about the $l=0$ and $l=1$ resonances of the individual
spheres. From Fig.~\ref{fig6}, we expect a single $l=0$ resonance
band of $\Lambda_1$ symmetry about $\omega S/c_a\approx2.2$; and
three $l=1$ bands about $\omega S/c_a\approx3.9$, of which one
should be non-degenerate ($\Lambda_1$ symmetry) and one doubly
degenerate ($\Lambda_3$ symmetry). We expect a degree of
hybridization between bands of the same ($\Lambda_1$) symmetry,
and this naturally leads to the bands shown in Fig.~\ref{fig7}(a),
except for the one flat band, a doubly degenerate band of
$\Lambda_3$ symmetry, about $\omega S/c_a=3.426$, which appears
to have a very different physical origin (see below).

In the long-wavelength limit ($\omega \rightarrow 0$) we obtain a
linear dispersion curve, the slope of which gives an effective
velocity of sound in the composite medium, $\bar{c}=278$ m/s,
which is in very good agreement with result, $\bar{c}=291$ m/s, of
the effective-medium approximation.~\cite{Gaunaurd} We note that
this agreement is greatly improved at lower values of the
volume-filling fraction $f$. It is worth noting, also, that the
effective velocity of sound in a system of silica spheres in air
is smaller than in air because of the high density contrast
between silica and air. Indeed, in the limit $\rho_a << \rho_s$,
the effective-medium approximation gives practically identical
results with the simple expression $\bar{c}/c_a=[2/(2+f)]^{1/2}$,
which gives $\bar{c}<c_a$ at any value of $f$.

\begin{figure}
\centerline{\includegraphics*[height=5cm]{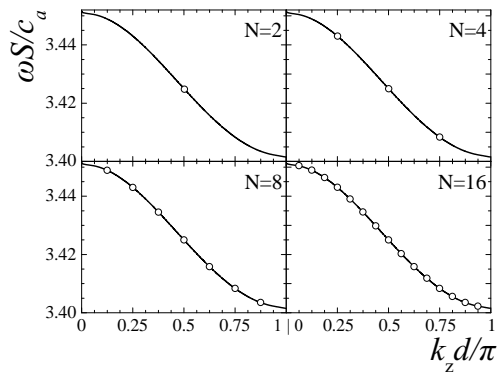}} \caption{The
deaf band about $\omega S/c_a=3.426$ of Fig.~\ref{fig7}(a) (solid
line). The doubly degenerate eigenfrequencies of the
corresponding bound states in slabs of $N=2,4,8,16$ planes of
spheres are noted by circles.}\label{fig8}
\end{figure}

It seems that the modes corresponding to the flat band about
$\omega S/c_a=3.426$, which are deaf modes, are those of a
wavefield highly concentrated between consecutive planes of
spheres, with very little interaction between neighbor regions of
high concentration. In order to demonstrate the above, we looked
for the eigenmodes of the acoustic field, for ${\bf k}_{\|}={\bf
0}$, in a slab of $N=2,4,8,16$ planes of spheres, in the manner
described for the corresponding problem of the electromagnetic
field,~\cite{Karathanos} over a frequency range about the said
flat band of Fig.~\ref{fig7}(a). Because of the two-dimensional
translation symmetry of the slab, the wavefield, in the air
regions, can be expanded [as in Eq.~(\ref{eq:unn+1})] in a series
of plane waves with wavevectors ${\bf K}_{\bf g}^{\pm}=\left({\bf
g}, \pm \left[(\omega/c_a)^2-{\bf g}\right]^{1/2}\right)$; now,
as it turns out, for the eigenmodes in question the term
corresponding to ${\bf g}={\bf 0}$ in the above plane-wave
expansion vanishes, which implies that the wavefield decreases
exponentially on both sides out of the slab, since $\omega/c_a <
|{\bf g}|$ for ${\bf g}\neq{\bf 0}$. We can say that these
eigenmodes correspond to bound (in the $z$ direction) states of
the acoustic field. Our results are shown in
Fig.~\ref{fig8}. For
a single pair of planes ($N=2$) we obtain a pair of such modes
corresponding to a doubly degenerate eigenfrequency. For three
pairs of planes ($N=4$) we obtain three pairs of doubly
degenerate eigenfrequencies, and similarly in each case we obtain
a pair of doubly degenerate eigenfrequencies for every pair of
planes in the slab. The eigenfrequencies corresponding to the
bound states of the different slabs in Fig.~\ref{fig8}, have been
plotted against values of the reduced wavenumber $k_{zi}=\pi
i/Nd$, $i=1,2,\ldots N-1$, for the slab consisting of $N$ planes,
to show how this band of states develops with increasing $N$. In
every case, $d$ equals the separation of two consecutive fcc
(111) planes in the infinite crystal, so that $Nd$ equals the
thickness of the slab. We note that the eigenfrequencies of the
finite slabs coincide with the corresponding eigenfrequencies on
the dispersion curves (solid lines in Fig.~\ref{fig8}) of the
infinite crystal. The question now arises as to whether one can
see experimentally this band. The corresponding modes do not
couple to an incident wave. This is demonstrated quite clearly in
Fig.~\ref{fig7}(b), which shows the transmission coefficient of a
wave incident normally on a slab of the material consisting of 32
planes of spheres. We can see that the transmission coefficient
vanishes at and about $\omega S/c_a=3.426$ where the band under
consideration exists. However, we have verified that the said
band survives for ${\bf k}_{\|}\neq{\bf 0}$ (at least in the
neighborhood of ${\bf k}_{\|}={\bf 0}$) where it couples with an
incident wave of the same ${\bf k}_{\|}$ leading to measurable
transmittance.

\begin{figure}
\centerline{\includegraphics*[height=5cm]{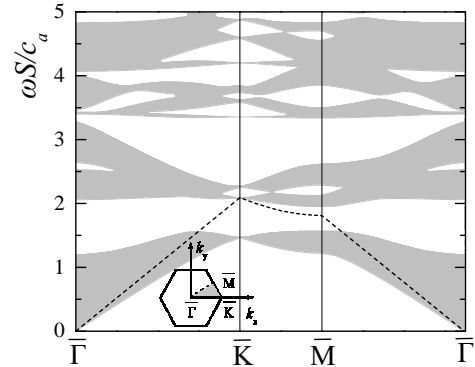}}
\caption{Projection of the phononic frequency band structure of a
fcc crystal of silica spheres in air, with $f=74\%$, on the SBZ
of the fcc (111) surface, along the symmetry lines shown in the
inset. Propagating waves in the air about a slab of the crystal
exist for frequencies above a threshold value (a function of ${\bf
k}_{\|}$) $\omega_{inf}=c_a |{\bf k}_{\|}|$ denoted by the dashed
line.}\label{fig9}
\end{figure}

In Fig.~\ref{fig9}, we present the projection of the frequency
band structure of the acoustic field of the phononic crystal
under consideration, on the symmetry lines of the SBZ of the fcc
(111) surface. The shaded regions extend over the frequency bands
of the acoustic field: at any one frequency within a shaded
region, for a given ${\bf k}_{\|}$, there exists at least one
propagating acoustic mode in the infinite crystal. The blank areas
correspond to frequency gaps. We note that knowing the modes with
${\bf k}_{\|}$ in the shaded area ($\bar{\Gamma}\, \bar{\rm K}\,
\bar{\rm M}$) of the SBZ and $-\pi/d < k_z \leq\pi/d$ is
sufficient for a complete description of all the modes in the
infinite crystal. The modes in the remaining of the reduced ${\bf
k}$ space are obtained through symmetry. One clearly sees that for
the opal under consideration, and unlike the colloidal crystal
studied in Section~\ref{poly}, one obtains an omnidirectional
frequency gap extending from $\omega S/c_a=1.59$ to $\omega
S/c_a=1.95$. We verified that this is indeed so by calculating
the band structure at a sufficient number of ${\bf k}_{\|}$
points in the SBZ. Finally, we should note the existence of a
very narrow omnidirectional gap at higher frequencies, extending
from $\omega S/c_a=3.30$ to $\omega S/c_a=3.34$.
\newline

 This work has been supported by the Institute of Communication
 and Computer Systems (ICCS) of the National Technical University
 of Athens. Partial support from the University of
 Athens is also acknowledged. R.~Sainidou is supported by the
 State Foundation (I.K.Y.) of Greece.
{}

\end{document}